\documentclass[]{aastex631}
\usepackage{amsmath}
\usepackage{amsfonts}
\usepackage{amssymb}
\usepackage{natbib}
\usepackage{cases}

\usepackage{color}
\newcommand{\af}[1]{\textcolor{black}{#1}} 

\shortauthors{Guinard et al.}

\begin{document}
\title{
Coupled geophysical and thermal constraints link between Mars basal molten layer  and  the planet viscosity profile\\
}

\correspondingauthor{A. Guinard, A. Fienga}
\email{alex.guinard@etu.univ-cotedazur.fr, agnes.fienga@oca.eu}

\author{Alex Guinard}
"\affiliation{G\'eoazur, CNRS, Observatoire de la C\^ote d'Azur, Universit\'e C\^ote d'Azur, Valbonne, France}
\author{Agnès Fienga}
\affiliation{G\'eoazur, CNRS, Observatoire de la C\^ote d'Azur, Universit\'e C\^ote d'Azur, Valbonne, France}
\author{Anthony M\'emin}
\affiliation{G\'eoazur, CNRS, Observatoire de la C\^ote d'Azur, Universit\'e C\^ote d'Azur, Valbonne, France}
\author{Cl\'ement Ganino}
\affiliation{G\'eoazur, CNRS, Observatoire de la C\^ote d'Azur, Universit\'e C\^ote d'Azur, Valbonne, France}



\date{}
\begin{abstract}
Computing the tidal deformations of Mars, we explored various Mars internal structures by examining profiles that include or exclude a basal molten layer within the mantle and a solid inner core. By assessing their compatibility with a diverse set of geophysical observations we show that despite the very short periods of excitation, tidal deformation is very efficient to constraint the Mars interior. We calculated densities and thicknesses for Martian lithosphere, mantle, {core-mantle boundary} layer and core and {found them} coherent with preexisting results from other methods. We also estimated new viscosities for these layers. We demonstrated that the geodetic record associated with thermal constraint is very sensitive to the {presence} of a basal molten layer in deep Martian mantle, and less sensitive to the solid core. Our results also indicate that the existence of the basal molten layer necessarily comes together with an inversion of viscosity between the lithosphere and the mantle. In this case, we could attribute this reverse viscosity contrast to the poor hydration state of Martian mantle and we underlined that this result prevents a strict Earth-like plate tectonics on Mars. The existence of the basal molten layer is also associated with a non-inversion of viscosity between the core-mantle boundary layer and the liquid core. Finally, in our results, a basal molten layer is incompatible with the existence of a solid inner core. Efforts to detect basal molten layer are then of prime importance to decipher the Martian interior. {Inversely, viscosity profiles appear to be very good tools for probing the existence of such molten layer at the base of the Mars mantle.}
\end{abstract}


\section{Introduction}
\label{sec:introduction}


The marsquake S1000a, detected by NASA's InSight seismometer \citep{banerdt2020initial}, represents a significant breakthrough, marking the first observation of a P-wave diffracted (hereafter labelled Pdiff) along \af{the layer in between the lower mantle and the core also called} the core-mantle boundary \af{layer} (CMB) \citep{horleston2022far}. On one hand, the majority of existing profiles, such as those presented in studies by \cite{drilleau2022marsquake} or \cite{duran2022seismology}, assume a mantle \af{homogeneous in composition}. However, these profiles show  fits \af{not matching or poorly matching} with the differential measurements of arrival times between Pdiff and PP (P-wave reflected once at the planet's surface). On the other hand, alternative profiles, like the one proposed by \cite{duran2022observation}, suggest the existence of a heterogeneous mantle. According to \cite{samuel2022testing}, these profiles require a significant reduction in velocity in the deep mantle to consistently explain the observations. Indeed, recent publications by \cite{samuel23} and \cite{khan2023evidence} highlight a thin layer of molten silicate above the core of Mars. This layer is believed to result from the solidification of a primitive magma ocean, forming an iron-enriched basal layer containing heat-producing elements \citep[e.g.][]{elkins2003magma, zeff2019fractional}.\\

Tidal forcing exerted on Mars induces a global long wavelength deformation of the
planet. Due to the periodic nature of the orbit of planetary bodies,
the induced tidal deformation on Mars is also periodic but with a slight delay that witnesses
the anelastic tidal dissipation of the planet. As the tidal deformation
and dissipation depend on the inner structure of Mars, it is also possible
to probe Mars interior from tidal observations. Analysis of radio-tracking monitoring
of orbiters have revealed the tidal change of the Martian gravity field \citep{konopliv2020detection}.
More specifically the non-dimensional tidal Love number k$_2$ relating
linearly the degree-2 perturbation of the gravity potential to the forcing potential, has been estimated.
\af{Furthermore, by considering the orbital interactions between Mars and its satellite Phobos, it was possible to detect the tidal dissipation induced by Phobos on Mars and to evaluate the tidal quality factor at the Phobos excitation frequency \citep{pou2022tidal}.}
\af{In combining these two information, it is then possible to quantify the amplitude and the lag of the change in the gravity potential.}\\

In this paper, we explore various Mars internal structures by examining profiles that include or exclude a solid inner core. Simultaneously, we aim to detect the presence of \af{a} basal molten layer (BML) within the mantle by assessing their compatibility with a diverse set of geophysical observations \af{and thermal constraints}. Sec. \ref{sec:method} outlines our semi-analytical approach for estimating tidal \af{complex} Love numbers for each Mars internal structure. Profiles consistent with observations are identified according to the procedure outlined in Sec. \ref{sec:method}. The results of this statistical selection are presented in Sec. \ref{sec:results}, where we also assess the compatibility of our profiles with existing \af{thermal constraints} to propose \af{possible} inner structures, the viscosity profiles being the most stringent aspect of this selection. In Sec. \ref{sec:discussion}, we discuss the implications of these results and how tidal deformation, mass, moment of inertia and \af{thermal constraints} can help to decipher the Martian interior.

\section{{Tidal deformation and internal structure}}
\label{sec:method}

\subsection{Computing tidal deformation parameters}

For a viscoelastic planetary body, the Love number k$_2$ depends on the period of the tidal forcing and the interior structure of the body. 
We use the software ALMA (plAnetary Love nuMbers cAlculator) initially developed to compute Love numbers for a body subject to Heaviside time-history loading with applications to  glacial isostatic adjustment studies \citep{spada2006using, spada2008alma}. With ALMA$^3$, it has recently been extended to compute 
periodic tidal Love numbers \citep{2022GeoJI.231.1502M}. This software numerically integrates the gravito-viscoelasticity equations in the frequency domain 
and subsequently employs a numerical Laplace inversion to 
retrieve Love numbers in the time domain 
using the Post-Widder formula \citep{post1930generalized,widder1934inversion,widder1941laplace}. This non-conventional strategy allows rapid and accurate 
computation of Love numbers for a spherically symmetric, 
non-rotating, incompressible and self-gravitating viscoelastic 
body. 
As a consequence, ALMA requires as inputs the 
parameters that describe the multi-layered 1-D rheological profile, namely radius, density, rigidity and viscosity for each layer. It also requires 
the frequency $\omega = \frac{2\pi}{T_f}$ of the periodic excitation $e^{i\omega t}$, where t is time and $T_f$ is the period of the tidal forcing. In the frequency-domain, $k_2$ is complex. Due to energy dissipation, 
the response of the planet is delayed. This implies that the time-dependant $k_2$ 
follows the relation \citep[][]{2022GeoJI.231.1502M}
\begin{equation}
k_2(t) = \| k_2(\omega) \| e^{i(\omega t-\epsilon)}
\end{equation}
where $\epsilon$ is the phase lag that 
can be estimated with  
\begin{equation}
\textrm{tan} (\epsilon) = - \frac{\textrm{Im}(k_2(\omega))}{\textrm{Re}(k_2(\omega))}.
\end{equation}
The ratio between the energy loss by dissipation  and the total energy of the body is often given in term of quality factor $Q$ which also depends on the excitation frequency $\omega$ such as:
\begin{equation}
Q(\omega) = - \frac{\| k_2(\omega) \| }{\textrm{Im}(k_2(\omega))}
\end{equation}

\subsection{General inversion strategy}


With the objective to use the tidal deformation as a tool to explore 
the inner structure of Mars we employed an inversion strategy as described by \cite{briaud2023constraints, briaud2023lunar} for the Moon and \cite{Saliby2023} for Venus. 
It consists in varying, using the \af{random walk exploration} method, the parameters required to estimate the tidal deformation. 
First, we use the observational  constraints given in Table \ref{tab:geodobs} to create a database containing profiles that satisfy the total mass and radius of Mars as well as the moment of inertia (see Table \ref{tab:exploration_range}). 
We then use ALMA to estimate the Love number $k_2(\omega)$ and  the dissipation $Q(\omega)$ of Mars, at 
the period of Phobos tidal forcing. 
We subsequently compare these two estimated parameters to those derived from the observations (Table \ref{tab:geodobs})  computing the $\chi^2$ for each model from the database. 
The results of this selection are presented in Table \ref{tab:sol_andrade_5&6l} where we use the 3-$\sigma$ of the $\chi^2$ distribution as a criterion for selecting profiles compatible with the observations.

\subsection{Tidal Observations}

\af{Geophysical Mars observations used as constraints for our inversion strategy are presented in Table \ref{tab:geodobs} and correspond to} : the mean radius R  \citep{seidelmann2002}, the total mass M \citep{konopliv2016improved}, the normalized moment of inertia C/M$R^2$ \citep{bagheri2019tidal}, the tidal Love number $k_2$ \citep{konopliv2020detection} and the quality factor Q \citep{pou2022tidal}. 
Besides, \cite{pou2022tidal} showed that the 
difference between $k_2$ predicted at the excitation period of Phobos (5.55\~h) and that estimated from observations at the excitation period of the Sun (12.32\~h) is significantly lower 
than the error on the latter. It 
is also reasonable to assume that the observational uncertainties on these 
two values are of similar magnitude. Thus, as suggested by \cite{pou2022tidal}, we consider $k_2$ and Q 
to be that induced by the Phobos excitation period. 
More about the Mars tidal parameters can be found in the review by 
\cite{Bagheri22}.



\begin{table}
    \centering
    \small
    \caption{{Data used to constrain the internal structure of Mars}}
    \begin{tabular}{l l l l l}
    \hline
    Data & Symbol & Value & $3-\sigma$ & Reference \\
    \hline
    Mean radius (km) & R & $3389.5$ & $\pm0.6$  & \cite{seidelmann2002} \\
    Total mass (kg) & M & $6.417\times10^{23}$ & $\pm8.943\times10^{19}$ & \cite{konopliv2016improved} \\
    Moment of Inertia & $C/MR^2$ & $0.363779$ & $\pm0.0003$ & \cite{bagheri2019tidal} \\
    Tidal Love number & $k_2$ & $0.174$ & $\pm0.024$ & \cite{konopliv2020detection} \\
    Dissipation & $Q$ & $93$ & $\pm25.2$ & \cite{pou2022tidal} \\
    Phobos period (h) & $T_S$ & 5.55 &   &   \\
    \hline
    \end{tabular}
    \label{tab:geodobs}
\end{table}

\subsection{Profile configuration}
\label{sec:profile}

To calculate the viscoelastic tidal Love numbers $k_2$ using ALMA, we need to set up 1D profiles that describe the hypothetical interior structure of Mars. The input parameters for ALMA are the radius, the density, the rigidity, the viscosity and the rheology for each layer. 
We specifically consider 5-layer and 6-layer Mars profiles. Both type of profiles have 
a crust, a lithosphere, a mantle and a mantle basal layer. The difference is in the structure of the core which is uniform for the 5-layer profiles and differentiated for the 6-layer profiles. Hereafter we describe each layer and Table \ref{tab:exploration_range} shows the structure of the 1D profiles that we have considered and the exploration ranges of the parameters assumed for each layer.



\paragraph{The crust }
The seismic studies of \cite{kim2023global} and \cite{li2022constraints} have allowed to give strong constraints on the Martian crust. Indeed, by combining seismology and gravimetry, \cite{kim2023global} managed to estimate the average thickness at 49$\pm$7~km of the Martian crust thanks to the largest marsquake (Mw = 4.6) detected by the InSight mission. \cite{li2022constraints} used the data from the farthest marsquake from the lander and concluded that the crust would be globally composed of two sub-layers with different propagation speeds rather than three as under the landing site. The choice of density is based on the study of \cite{wieczorek2022insight} who evaluated three density profiles among which we will use the values of the crust with a uniform density equal to 2.9~g/$cm^3$. We use the elastic rheology for this layer.

\paragraph{The lithosphere and the mantle }
The input parameters concerning the lithosphere and the mantle are based on the work of \cite{khan2021upper} and \cite{drilleau2022marsquake}. These studies show that Mars has a much thicker lithosphere than that of the Earth, about 500~km, which implies a reduction in the size of the mantle. Indeed, their thermo-chemical studies 
have brought strong constraints on the thicknesses of the lithosphere and the mantle. The velocity profiles of \cite{drilleau2022marsquake} revealed a slowing down of the S waves in this layer sometimes called LVZ (low velocity zone) in the literature. We consider 
the density of the lithosphere by exploring a 
range of values that 
are similar to the density profile of \cite{plesa2021seismic}. The density of the mantle is deducted from the other layers to conserve the total mass of the planet. To follow the studies carried out by \cite{bagheri2019tidal}, we use the Andrade rheology for the lithosphere and the mantle.

\paragraph{The CMB and the uniform core }\label{paragraph2.2.2.3}
By combining the study of the travel time of P waves diffracted at the core-mantle interface and the geodynamic constraints on the composition of the core, 
\cite{khan2023dlayer} 
have identified a layer between the core and the mantle called CMB composed mainly of molten silicate. 
This study 
estimates this layer is 145$\pm$25~km thick. 
This thickness implies a reduction in the radius of the core to 1690$\pm$50~km instead of 1830$\pm$40~km as recently estimated by \cite{stahler2021seismic}. Conversely, still according to \cite{khan2023evidence}, the density of the core is constrained to increase to 6.65$\pm$0.15~g/$cm^3$. We choose a wide range of density values for the CMB in order to explore a maximum of possibilities. Due to the lack of knowledge about CMB and core viscosities we explore a wide range of values. According to the last shear wave velocity profiles \citep[e.g.][]{huang2022seismic}, there seems to be no S-wave at the level of the CMB, thus we consider this layer as fluid.  We use the Newton rheology for the CMB and the uniform core.

\paragraph{The outer and inner core }
This paragraph only concerns the case of a 6-layer model, i.e. a model in which the core is subdivided in two parts: a liquid outer core and a solid inner core. The work of \cite{hemingway2021history} aims at showing that the absence of a magnetic field on Mars does not necessarily imply the absence of a solid inner core. Indeed, thanks to simulations of thermal evolution of the planet, the study was able to demonstrate that the presence of a solid inner core is possible and this implies that the Martian dynamo will reactivate in the future as it has previously been able to reactivate for a short period according to the work of \cite{lillis2005evidence}. Their simulation shows that the solid inner core could be about 400~km in radius and reach 7.7~g/$cm^3$ in density. Moreover, recent seismic studies by \cite{irving2023first} show that if a solid inner core exists then its radius does not exceed 750~km. We choose to explore a region around the values given by \cite{khan2023evidence} in the previous paragraph for the radius and the density of the outer core. 
For the inner core, we impose a maximum radius of 750$\pm$50~km to follow the condition given by \cite{irving2023first}. 
We use exploration intervals wide enough not to miss profiles compatible with the observations. We assume that the rheology of the inner core is elastic and that of the outer core is Newtonian.

\begin{table}
    \centering
    \small
    \caption{{General parameters of 1D profiles for the interior of Mars.} Values in square brackets [ ] indicate the range of parameters that vary randomly and uniformly between profiles. The stars * represent the values that are inferred from the other layers to preserve the size and total mass of Mars. The inner core layer is only used for the 6-layer profiles. For Andrade rheology $\alpha=1/3$.}
    \begin{tabular}{c|c c c c c c c}
    \hline
    Layer & &Crust & Lithosphere & Mantle & CMB & Core/Outer Core & Inner Core \\
     & & Cr & Li & Ma & CMB & LC& SIC \\
    \hline
     & && & & & & \\
    Radius & km & 3389.5 & [3319: 3362] & * & [1700: 2000] & [0: 1800] & [0: 800]\\
     & && & & & & \\
    Density & g/$cm^3$& [2.7: 3.1] & [3: 3.8] & * & [3: 10] & [4: 10] & [4: 10]\\
     && & & & & & \\
    Rigidity &GPa& [20: 25] & [20: 70] & [20: 130] & - & - & - \\
     & && & & & & \\
    Viscosity & $log_{10}$(Pa.s)  & - & [14: 30] & [18: 30] & [1: 30] & [1: 30] & - \\
    & & & & & & & \\
    Rheology & - & Elastic & Andrade & Andrade & Newton & Newton & Elastic \\
    & & & & & & & \\
    \hline
    \end{tabular}
    \label{tab:exploration_range}
\end{table}

\subsection{Viscosity profile sensitivity}
\label{sec:sensitivity}

\begin{figure}[!ht]
    \centering
    \includegraphics[width=1\textwidth]{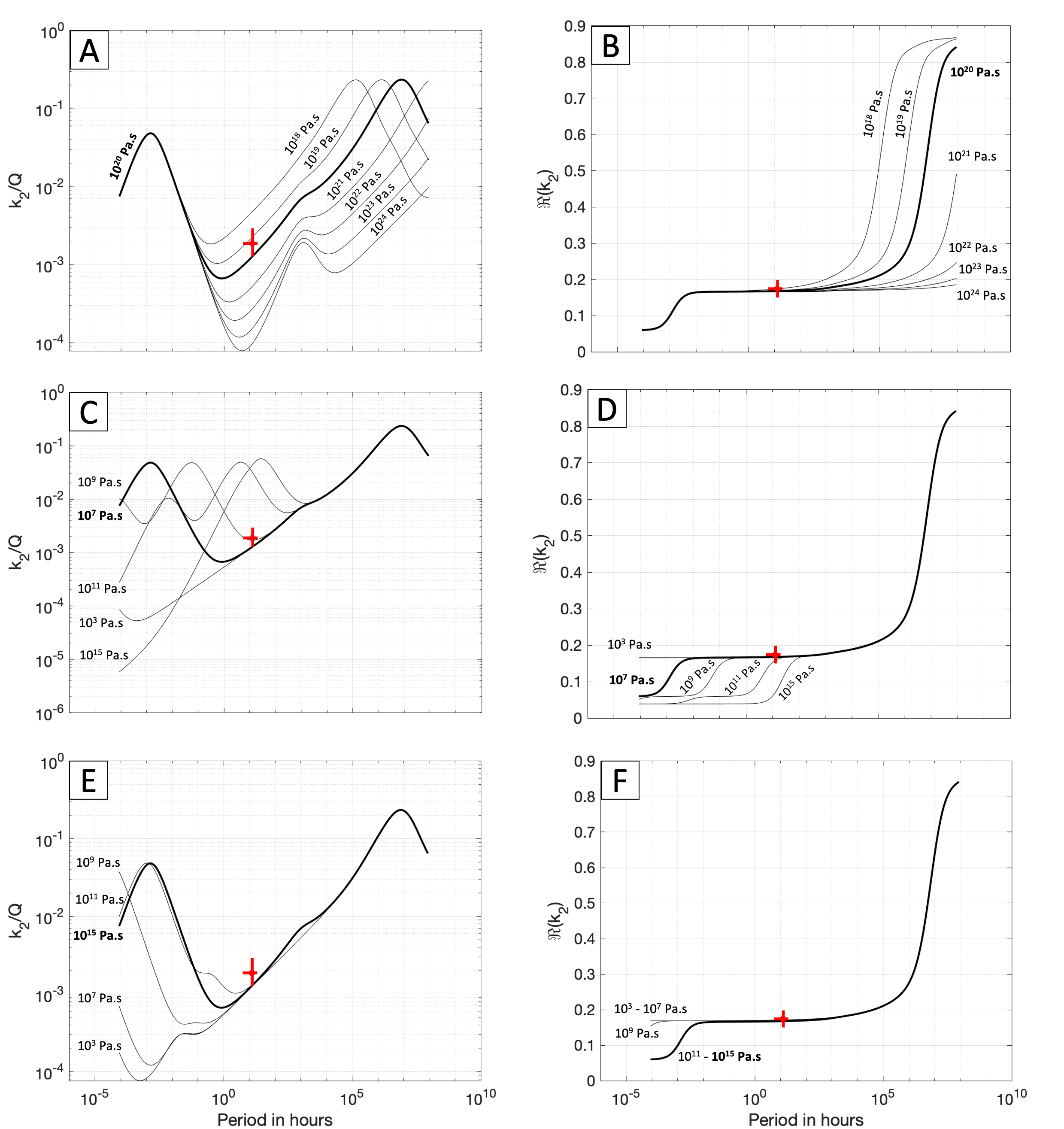}
    \caption{{Behavior of the $k_2$/$Q$ ratio and of the real part of $k_2$ as a function of the period.} Each curve corresponds to a different mantle (A-B), CMB (C-D) or core (E-F) viscosity. In bold, the reference model considered and in red, the value and the error bars of $k_2$/$Q$ and of $\Re{(k_2)}$ corresponding to the period of excitation induced by Phobos at 5.55~h. The error bars in x-axis correspond to the values of \cite{pou2022tidal}.}
    \label{fig:sensitivity}
\end{figure}

The role of the viscosity of each layer is of primary importance when assessing the tidal dissipation and the tidal response delay of a planet. Hence we investigate the contribution  of the mantle, the CMB and the core in 
dissipating energy through tidal deformation and in bringing constraints on the viscosity structure of Mars. 
Here, we use as a reference model for the interior of Mars, \af{one possible} 5-layer model \af{meeting geophysical constraints} and with the mean radius, density and rigidity defined in Table \ref{tab:sol_andrade_5&6l}. 
\af{In order to assess the sensitivity of the tidal deformation (considering the imaginary and real parts of the $k_2$ Love number estimated with ALMA) to the viscoelastic behaviour of the planet even for very rapid (few hours) excitation periods, we considered different possible viscosities for the lithosphere, the mantle, the CMB and the core respectively, starting with} reference values such as 10$^{24}$, 10$^{20}$, 10$^{7}$, 10$^{15}$~Pa.s for the lithosphere, the mantle, the CMB and the core respectively. 
We then plot the ratio $\| k_2(\omega) \|/Q$ and the value $\textrm{Re}(k_2(\omega))$ computed 
for periods ranging from  10$^{-4}$ to 10$^{8}$~hours using the reference model and varying the viscosity of 
the mantle (Fig.  \ref{fig:sensitivity}-A and \ref{fig:sensitivity}-B)
the CMB (Fig.  \ref{fig:sensitivity}-C and \ref{fig:sensitivity}-D) and 
the core (Fig.  \ref{fig:sensitivity}-E and \ref{fig:sensitivity}-F). 
We obtain that at the tidal excitation period of Phobos,  $\textrm{Re}(k_2(\omega))$ cannot discriminate the viscosity of the mantle, the CMB and the core 
over the range 10$^{18}$-10$^{24}$, 10$^{3}$-10$^{11}$ and 
10$^{3}$-10$^{15}$~Pa.s, respectively. 
The same is obtained for $\| k_2(\omega) \|/Q$ regarding 
the viscosity of the core. However, $\| k_2(\omega) \|/Q$ 
varies significantly depending on the viscosity of 
the mantle and the CMB. Consequently it appears to be  
a good parameter to constrain the viscosity of 
\af{these two layers}.

\begin{table}[]
    \centering
        \caption{Radii (given at the top of the layer), densities, viscosities and rigidities of profiles selected with the method described in Sec. \ref{sec:method}. The values presented in this table represent the 0.5, 0.995 and 0.005 quantiles. Crust (Cr), Lithosphere (Li), Mantle (Ma), Core-Mantle Boundary layer (CMB), Fluid (or liquid) Core (for 5-layer profiles)/ Fluid (or liquid) Outer Core (for 6-layer profiles)(\af{LC}),  Solid Inner Core (SIC) (for 6-layer profiles).}
    \begin{tabular}{c| l c c c c c c}
    Number of profiles & & \multicolumn{1}{c}{Cr} & \multicolumn{1}{c}{Li} & \multicolumn{1}{c}{Ma} & \multicolumn{1}{c}{CMB} & \multicolumn{1}{c}{\af{LC}} & \multicolumn{1}{c}{SIC} \\
    \hline
     2,971 &  {\bf{5-layer profiles}}  & & &  & & \\
    &   Radius [km] & 3,389.5 & 3,341$^{3,361}_{3,320}$ & 2,886$^{3,343}_{2,409}$ & 1,786$^{1,899}_{1,701}$ & 1,641$^{1,770}_{1,540}$ & -\\
     &  Density [g.cm$^{-3}$] & 2.90$^{3.10}_{2.70}$ & 3.35$^{3.52}_{3.01}$ & 3.69$^{4.23}_{3.47}$ & 5.08$^{6.84}_{3.62}$ & 6.63$^{7.74}_{5.67}$ & -\\
    &   Viscosity [log$_{10}$(Pa.s)] & - & 20.09$^{29.92}_{14.18}$ & 21.16$^{29.95}_{18.80}$ & 7.14$^{13.81}_{1.05}$ & 12.78$^{29.76}_{1.13}$ & -\\
    & Rigidity [GPa] & 22.50$^{24.98}_{20.03}$ & 44.34$^{69.70}_{20.18}$ & 75.44$^{124.81}_{55.03}$ & - & - & -\\
    \\
     3,219 &  {\bf{6-layer profiles}}  & & &  & & \\
    &   Radius [km] & 3,389.5 & 3,340$^{3,361}_{3,320}$ & 2,885$^{3,336}_{2,410}$ & 1,792$^{1,899}_{1,701}$ & 1,646$^{1,766}_{1,537}$ & 463.7$^{796.1}_{5.0}$\\
     &  Density [g.cm$^{-3}$] & 2.90$^{3.10}_{2.70}$ & 3.34$^{3.53}_{3.01}$ & 3.71$^{4.27}_{3.47}$ & 5.08$^{6.67}_{3.62}$ & 6.53$^{7.68}_{5.44}$ & 8.32$^{9.98}_{6.09}$\\
    &   Viscosity [log$_{10}$(Pa.s)] & - & 22.07$^{29.92}_{14.10}$ & 24.06$^{29.94}_{18.03}$ & 7.56$^{15.75}_{1.06}$ & 13.34$^{29.73}_{1.16}$ & -\\
    & Rigidity [GPa] & 22.46$^{24.96}_{20.02}$ & 44.20$^{69.68}_{20.31}$ & 75.99$^{125.11}_{21.77}$ & - & - & -\\
    \hline
    \end{tabular}
\label{tab:sol_andrade_5&6l}
\end{table}

\section{Results}
\label{sec:results}

\subsection{Geophysical constraints}
\label{sec:results1}

\begin{figure}
    \centering
    \includegraphics[width=0.8\textwidth]{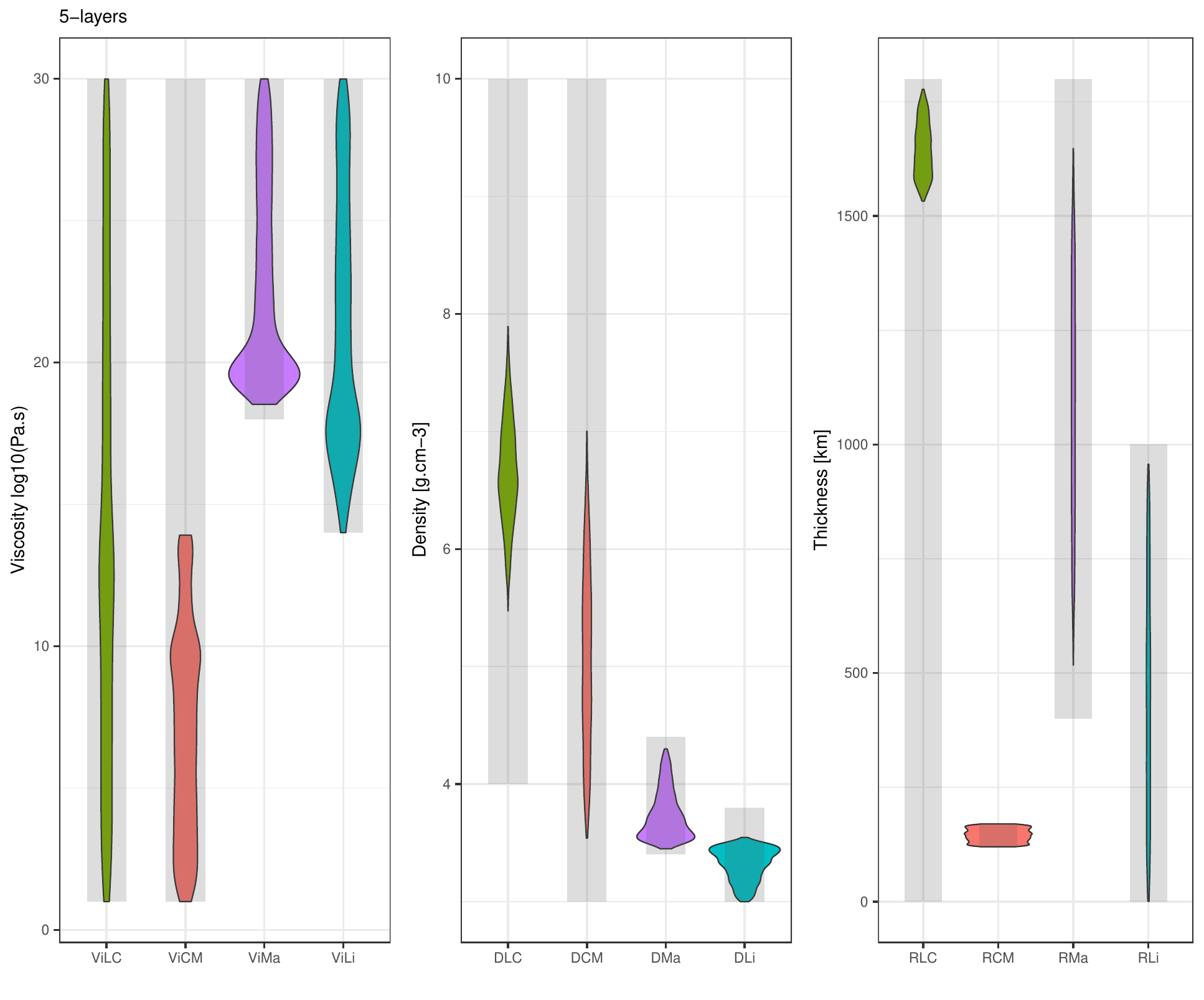}
    \caption{Distribution in viscosity, density and thickness for the 5-layer profiles presented in Table \ref{tab:sol_andrade_5&6l}. The gray bars in the background correspond to the exploration intervals of each parameter given in Table \ref{tab:exploration_range} before the $\chi^2$ filtering detailed in Sec. \ref{sec:method}.}
    \label{fig:violin_visco_5L_basic}
\end{figure}

As presented in Sec. \ref{sec:profile}, \af{we have considered} two types of profiles: one with a fixed crust, a lithosphere (Li), a mantle (Ma), a core-mantle boundary layer (CMB) and a fluid (or liquid) core \af{(LC)}(5-layer profiles) and with also a solid inner core (SIC) (6-layer profiles).
 Table \ref{tab:sol_andrade_5&6l} gives the main characteristics of the profiles selected \af{according to the mass, moment of inertia and tidal deformations (see Sec. \ref{sec:method}). We also plotted their distributions in viscosity, density and thickness on Fig. \ref{fig:violin_visco_5L_basic} for 5-layer profiles and Fig. \ref{fig:violin_visco_6L} for 6-layer profiles.} \\

\begin{figure}
    \centering
    \includegraphics[width=0.5\textwidth]{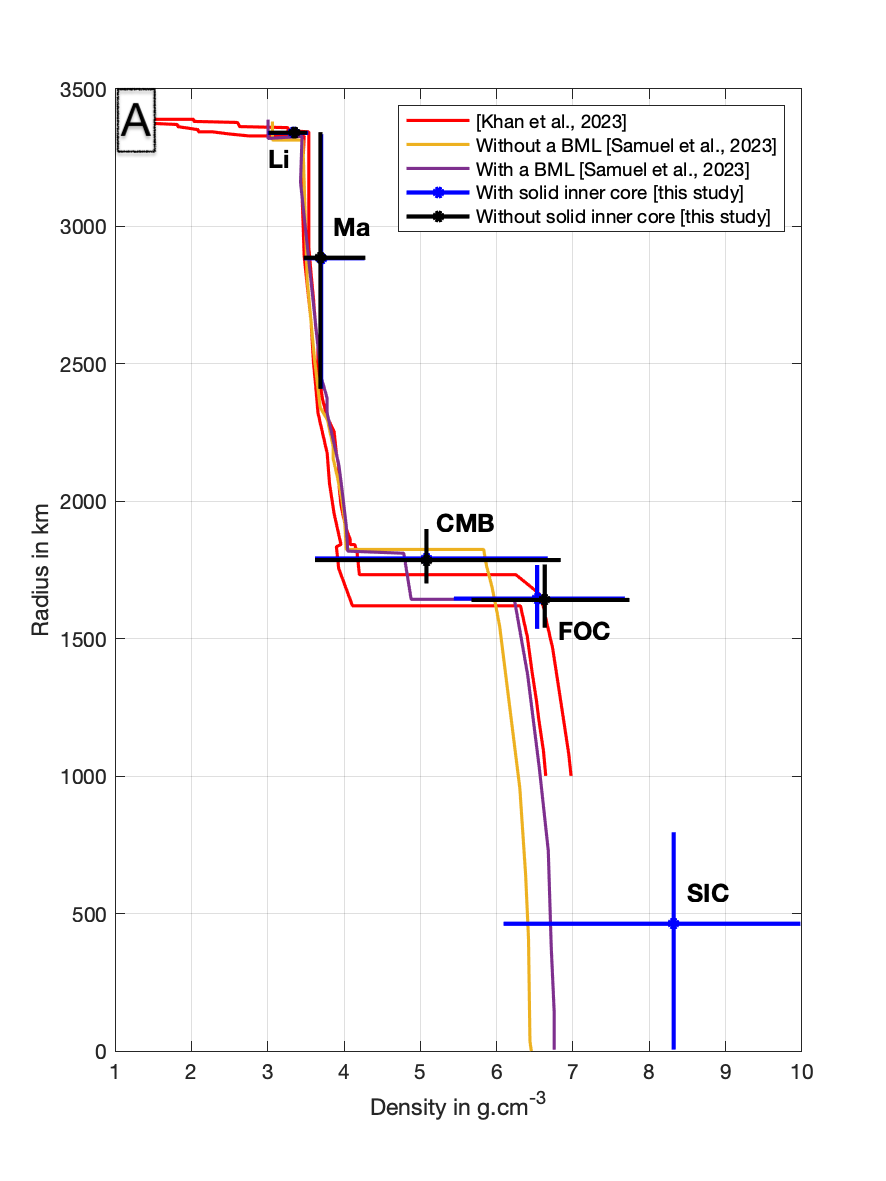}\includegraphics[width=0.5\textwidth]{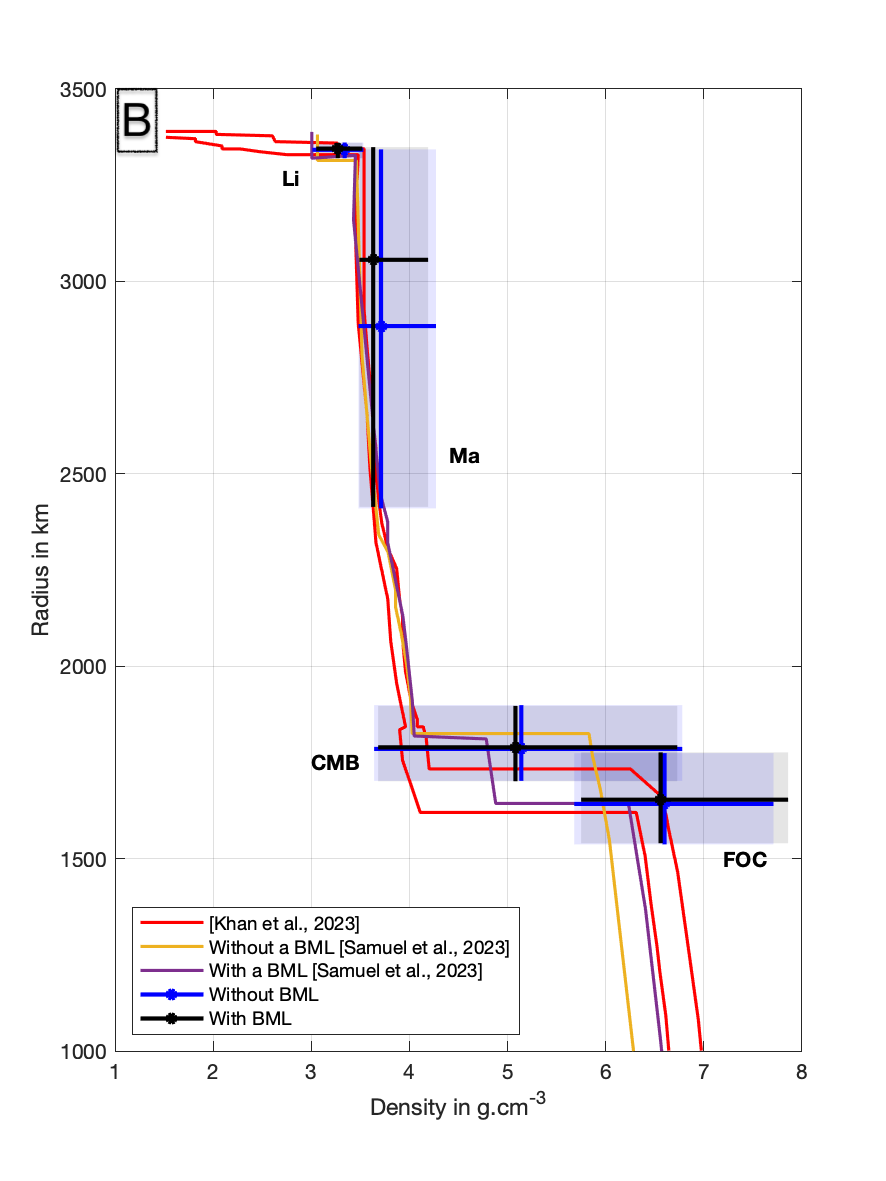}
    \caption{Density profiles from the profiles of the Tables \ref{tab:sol_andrade_5&6l} and \ref{tab:sol_LS} following the geophysical constraints (A) and both geophysical and thermal constraints (B). Each point is the median of the density distribution of the layer considered and the error bars are given by the quantiles 0.995 and 0.005. On Fig.B, The blue and black rectangles represent the acceptable depth/density surface given by the uncertainties. The red lines give the contour of the profile given by \cite{khan2023dlayer}, the orange line gives the density profile from \cite{samuel23} without BML and the purple one gives the \cite{samuel23} profile with BML.}
    \label{fig:densityBML}
\end{figure}

We first present here the main differences between the results for the 5 and 6 layer profiles.
Overall, the radii and density parameters remain stable, regardless of the choice between the 5 or 6-layer profiles. As one can see on Fig.  \ref{fig:densityBML}-A presenting density versus depth profiles, the radii and densities obtained for the crust and the mantle are in agreement with literature \citep[e.g.][]{khan2021upper,samuel23}. The CMB, for its part, is a thin layer of approximately 145 km thick and dense of 5.08 g.cm$^{-3}$ which seems to have a composition different from that of the mantle (approximately 3.70 g.cm$^{-3}$)  as suggested by \cite{khan2023evidence}. It is interesting to note that density compensation takes place exclusively between the fluid outer core (LC) and the solid inner core (SIC). Indeed, for the LC, the difference in density is very small (from 6.53 g.cm$^{-3}$ with SIC to 6.63 g.cm$^{-3}$ without SIC) due to the small volume occupied by the solid inner core (less than 0.5$\%$ of the total volume of the planet) and to the non-inversion of density between layers imposed to all the profiles. 
The results obtained with the 6-layer profiles specifically are presented in the Appendix \ref{sec:appendix}.

More generally, for both 5 and 6-layer profiles, the rigidity is not well constrained as they present uniform distributions which sweep the entire exploration interval. On Fig.  \ref{fig:violin_visco_5L_basic}, we can see how the geophysical constraints are very useful for reducing the intervals of possible thicknesses and densities. This is particularly true with the 5-layer profiles, for the LC density and thickness and for the density of the CMB. For the 6-layer profiles (see Fig.  \ref{fig:violin_visco_6L}) the constraints are less stringent.
Because we are considering viscoelastic rheologies of Mars and we use the quality factor Q obtained by \cite{pou2022tidal} as part of the profile selection, the viscosities of the different layers are an interesting outcome of the selection as it can be seen in Table \ref{tab:sol_andrade_5&6l} and Figs. \ref{fig:violin_visco_5L_basic} and \ref{fig:violin_visco_6L}. It is especially true regarding the viscosity of the CMB that has the interval of possible viscosity divided by a factor 2 after the selection, ending up with viscosity of the CMB systematically smaller than the one of the mantle and the one of the lithosphere. Furthermore, it also appears that it should be possible to have important inversion of viscosity between the CMB and the LC, despite a large distribution of possible CMB viscosities (from 10$^{1}$ to 10$^{13}$ Pa.s at 99.5$\%$ C.L.). Based on the results presented on Table \ref{tab:sol_andrade_5&6l} and Fig \ref{fig:violin_visco_5L_basic}, about 20$\%$ of the profiles may have a viscosity of the CMB at least 2 orders of magnitude bigger than the one of LC.  Such inversions are not expected and require more investigations and an improvement of the filtering. Additional filters have then to be considered such as thermal constraints. 
Finally, one can notice that the LC viscosity is not constrained as well as the one of the CMB. This is consistent with the weak sensitivity of the $k_2/Q$ ratio to the viscosity of the LC as indicated on Fig. \ref{fig:sensitivity} and commented in Sec. \ref{sec:sensitivity}.  

Most of the previous comments are also true for the 6-layer profiles (see Fig. \ref{fig:violin_visco_6L}).

\subsection{Thermal constraints}
\label{sec:thermal_constraint}

\subsubsection{Temperature estimation}
\label{sec:temp_prof}

\begin{table}
\caption{Initialization values extracted from \cite{samuel23}. The temperature of the top of the lithosphere is the average temperature taken at the bottom of the crust, for a thickness of the crust varying from 26 and 70 km. }
\centering
\begin{tabular}{c|c c}
     &  with BML & without BML\\
     \hline
    T lithosphere (K)  & \af{750} & 750 \\
    E$^{*}$ (kJ.mol$^{-1}$) & 120 $\pm$ 30  & 320 $\pm$ 70\\
    V$^{*}$ (cm$^3$.mol$^{-1}$) & 7.3 $\pm$ 2 & 4.6 $ \pm $ 2\\
    \hline
\end{tabular}
\label{tab:init}
\end{table}

We estimated the temperatures at the top of the mantle and at the top of the CMB in considering the viscosity contrasts between the lithosphere and the mantle and between the mantle and the CMB, with the hypothesis that all the layers are homogeneous. We use the equation for a convective mantle given by \cite{samuel23} such as:\\

\begin{equation}
    T(r) = T_L \times \frac{E^* + P(r)V^*}{E^* + P_LV^* + \ln{\left(\frac{\eta(r)}{\eta_L}\right)}R_gT_L}
    \label{eq:temp}
\end{equation}

with $R_g$ is the gas constant, $T_L$, $P_L$ $\eta_L$ are respectively the upper layer temperature, pressure and viscosity values, $P(r)$ and $\eta(r)$ are respectively the pressure and viscosity of the layer considered. $E^*$ and $V^*$ are given in the Table \ref{tab:init} and are the activation energy and volume values.
The temperature of the lithosphere (750K) is deduced from \cite{samuel23} as the average temperature at the top of the  lithosphere (bottom of the crust), corresponding to a crust thickness between 26 and 70 km.
With Eq. (\ref{eq:temp}) and the viscosities obtained in Table \ref{tab:sol_andrade_5&6l}, we computed the temperatures at the top of the mantle and at the top of the CMB.\\

Following \cite{samuel23}, we have used two set of values for the activation energy and volume: one set deduced from modeles including BML and one set without (see Table \ref{tab:init}).
{Finally, in order to estimate the impact of the enthalpy uncertainty, we vary the activation energy and volume using normal distributions based on uncertainties given by \cite{samuel23} and reported on Table \ref{tab:init}. We obtain about 1 million of estimations that are plotted in Fig.  \ref{fig:temp_profile1}.  
{By construction, these profiles match with the geophysical constraints described in Sec. \ref{sec:method}.}

\subsubsection{Solidus and Liquidus}
\label{sec:solidus}
Profiles can be filtered from the physical properties of putative BML. The pressure at BML depth is relatively well determined \citep{rivoldini2011geodesy} and this layer exhibits both liquid properties \citep[e.g.][]{samuel23} and solid characteristics from a seismological point of view \citep{khan2021upper}. Such properties are consistent with temperature of this layer being between solidus and liquidus. The chemical and mineralogical compositions of BML remain highly speculative but it is likely to have a mantle-like composition \citep{khan2018geophysical}. If Martian mantle does not significantly differ in composition from Earth mantle, the temperature range between solidus and liquidus at 190\,kbar is within 2100 and 2550\,K \citep{andrault2018deep}. It fixes possible range of temperature in deep Martian mantle as displayed in Fig.  \ref{fig:temp_profile1}. On the other hand, without BML, we consider that the bottom of the mantle is solid and the temperature is lower than 2300\,K.

\subsubsection{Temperature inversion at the mantle-lithosphere boundary}
\label{sec:inv}
An interesting feature of the temperature profiles presented in Fig.  \ref{fig:temp_profile1} is the temperatures at the top of the mantle. These temperatures have been obtained in considering a lithosphere at 750\,K as described in Sec. \ref{sec:temp_prof}. However, some of the produced profiles have a mantle temperature below the lithosphere temperature. Such an inversion of temperature is difficult to interpret because the evolution of the internal temperature with depth in Mars interior is the outcome of the competition between, on the one hand, residual heat and the heat produced by the radioactive decay of certain elements, primarily within the crust and mantle, and on the other hand, thermal dissipation through convection and conduction processes. Heat loss occurs through the surface, and as a result, the temperature within Mars necessarily increases with depth.
{In consequence, we decide to filter the profiles to those with a temperature of the mantle greater than the one of the lithosphere. The quantiles of the remaining profiles are represented with the white boxes of Fig.  \ref{fig:temp_profile1}.}
The distributions in viscosity, density and thickness for the profile selection considering both the solidus-liquidus constraints and the mantle-lithosphere non-inversion limits are presented in Fig.  \ref{fig:violon_all} for 5-layer profiles and in the appendix for 6-layer profiles (Fig.  \ref{fig:violon_6LwF}).

\section{Discussion}
\label{sec:discussion}
\subsection{A reverse lithosphere/mantle viscosity contrast related to a dry Martian mantle}
{On Fig.  \ref{fig:densityBML}-B, are plotted the density profiles deduced from our selection of 5-layer profiles with and without BML and considering both the geophysical and the thermal constraints. As it was already noticed 
when only the geophysical filter was applied, the addition of the thermal filter maintains the profiles in good agreement  with those from \cite{khan2023dlayer} and \cite{samuel23}.  }

The main impact of the thermal filtering is mostly on viscosity even if density intervals are also reduced.  This is particularly evident, on Fig.  \ref{fig:violon_all}, for the viscosity of the lithosphere and the one of the CMB, as well as for the density of the lithosphere and mantle. 

It is noteworthy that for 5-layer profiles with BML, the viscosity of the lithosphere is consistently approximately 2 orders of magnitude lower than the viscosity of mantle. This is contrary to what is documented for the Earth, where the viscosity of the lithosphere is 8–10 orders of magnitude larger than the viscosity of the \af{mantle} \citep{doglioni2011lithosphere}. The deformation of the major mineral of terrestrial planets mantle, olivine, largely depends on the temperature, and water content \citep{karato2088geodynamic}. With a geotherm quite similar between Earth and Mars, the significant difference in the viscosity of their \af{mantle} may indicate the extreme dryness of Martian mantle, as already proposed based on the chemical composition of Martian meteorites \citep{wanke1994chemistry}. Such reverse lithosphere/mantle viscosity profile would necessarily prevent the possibility of a strict Earth-like plate tectonics on Mars\\
On the other hand, without BML and with or without a solid inner core, such an inversion of viscosity between lithosphere and mantle is not present, and seems to indicate a profile closer to an Earth-like scenario.

\subsection{The presence of a BML prevents the existence of a solid core} 
\label{sec:nocorewBML}
On Fig.  \ref{fig:temp_profile1}, are plotted the temperatures versus depth profiles estimated following the procedures described in Sec. \ref{sec:thermal_constraint}. The black cross indicates the 0.5, 0.05 and 99.5 quantiles in temperatures and depths as given in Table \ref{tab:sol_andrade_5&6l} without thermal constraints.
The white cross gives the same but after filtering the profiles considering both the solidus-liquidus limit  at the CMB and the constraint of the non inversion of the constraint between the mantle and lithosphere.

\cite{helffrich17}, proposed that the temperature of the CMB is an indication of the existence of a liquid or solid core. Indeed, for $T_{CMB}<1300$\,K, the core is solid whereas for $T_{CMB}>1450$\,K the core is entirely liquid. Only  profiles without BML obtained after filtering on the thermal constraints given in Sec. \ref{sec:thermal_constraint}, follow the temperature conditions favoring the presence of a solid inner core (Figure \ref{fig:temp_profile1}). On the other hand, 6-layer profiles with BML that match the solidus-liquidus thermal constraints, have a temperature of the CMB (by filtering greater than 2100\,K) too high for meeting the CMB temperature requirement ($T_{CMB}<1300$K) for the existence of a solid inner core. We can then reject 6-layer profiles with BML and we detail the results obtained after thermal constraints in Table \ref{tab:sol_LS}.\\

\begin{figure}
    \centering
\includegraphics[scale=0.6]{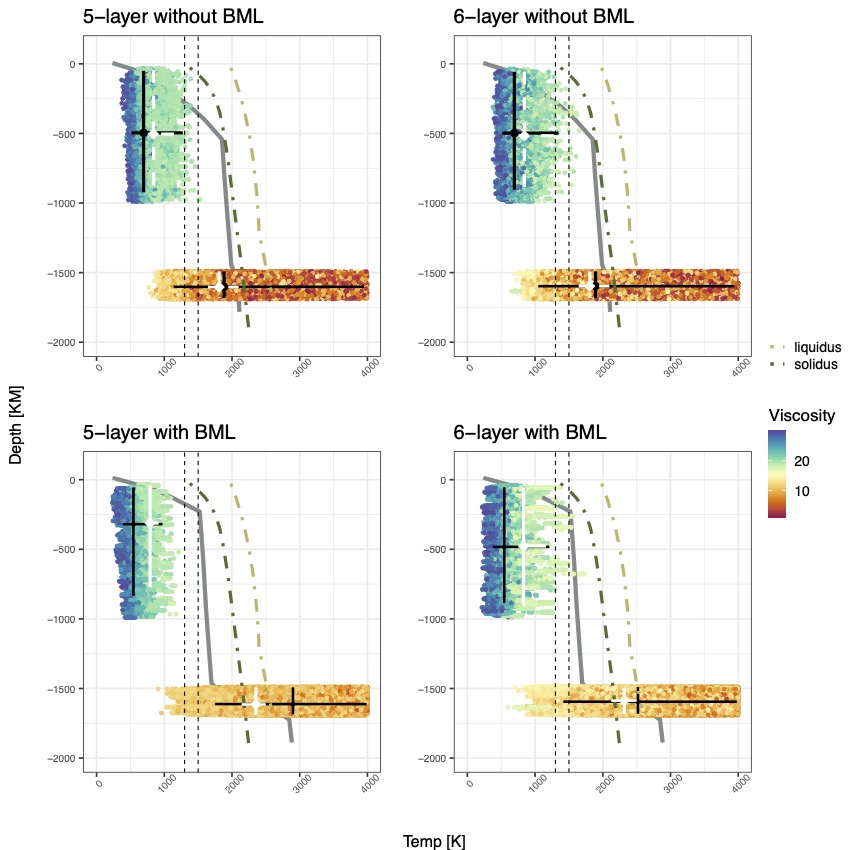}
    \caption{Temperature [K] versus Depth [km] profiles for different viscosities without inner core (A,B) and with inner core (C,D). Panels B and D gather profiles accounting for BML, whereas A and C present those  without BML. The black dots give the mean, 0.995 and 0.05 quantiles. The white dots give the same but in accouting for the liquidus and solidus constraints. The black dashed lines indicate the limit of cristallisation for the liquid core. Globally, the profiles with inner core tend to smaller temperatures, including those with BML. However, in this latest case, by accounting for the limit introduced by the liquidus and solidus thresholds (Sec. \ref{sec:solidus}), one can expect to have the temperature of the CMB in between 2300 and 2500K, which is imcompatible with the existence of a solid core. }
    \label{fig:temp_profile1}
\end{figure}

\subsection{Viscosity profiles and ratio of viscosities as  indicators of the BML}

On Fig.  \ref{fig:violon_all} are presented the new distributions of the viscosity, density and thickness of the 5-layer profiles respecting the geophysical and thermal constraints of Sec. \ref{sec:thermal_constraint}. 

It clearly appears that the existence of the BML imposes strong constraints particularly on the viscosities of the different layers. The intervals of viscosities for profiles with BML are indeed more reduced than the one without BML. For example, the lithosphere viscosity with BML can vary from greater than 10$^{16}$ to 10$^{22}$ Pa.s when it reachs up to  10$^{30}$ Pa.s without BML. 

At the bottom of the structure, the BML imposes a strong constraint for the liquid core with a viscosity less than  10$^{13}$ Pa.s, whereas without BML it can be as high as 10$^{30}$ Pa.s.

From these results, we can conclude that a profile associating a lithosphere with a viscosity greater than 10$^{16}$ Pa.s with a less than 10$^{13}$ Pa.s LC viscosity has a good probability to have a BML: less than 5$\%$ of the profiles without BML have viscosities in the limits of the profiles with BML.
This can also be easily visualized with the cumulative histograms of viscosity presented on Fig.  \ref{fig:proba}.  Are indicated in dashed lines, the quantiles obtained for profiles with BML at 10$\%$ limit of the CMB viscosity (in orange) and the 90$\%$ limit of the lithosphere viscosity (in black). For profiles without BML, more than 60$\%$ of the profiles have a viscosity of the CMB less than the BML-profiles inferior limit for the CMB and about 75$\%$ of the non-BML profiles have a lithosphere viscosity larger than the maximum value obtained for the BML-profiles. It is then clear that the BML imposes a very specific behaviour in the distribution of possible viscosities for all the layers, but more specifically for the CMB and the lithosphere. 

Additionally, on Fig.  \ref{fig:proba_ratio}, one can see than  100$\%$ of the BML-profiles have the viscosity of CMB greater than viscosity of liquid core and 90$\%$ have viscosity of lithosphere smaller than viscosity of mantle. For profiles with BML, the two conditions ($\eta_{CMB}> \eta_{LC}$ and  $\eta_{Ma}> \eta_{Li}$) represent 97$\%$ of the solutions whereas it gathers only 13$\%$ of profiles without BML. If, additionally, we consider the limits on the viscosity intervals imposed by BML ($\eta_{CMB}> 10^{10}$Pa.s and $\eta_{LC}< 10^{13}$Pa.s and $\eta_{Li}< 10^{22}$Pa.s), only 4$\%$ of the profiles without BML can match these requirements compared to 97$\%$ of the BML profiles.

Finally it is interesting to note that, the viscosity distributions of the 6-layer profiles without BML are very similar to the one of the 5-layer profiles without BML. This gives credits to the hypothesis that the BML is introducing more differences in the viscosity distributions than the introduction of the solid core.\\

\begin{table}[]
    \centering
        \caption{Radii (at the top of the layer), densities, viscosities and rigidities of the profiles presented in Table \ref{tab:sol_andrade_5&6l} for which the temperatures for the top of the mantle and the top of the CMB were calculated for 300 different values of $E^*$ and $V^*$ determined randomly from Gaussian manner in the ranges given in Table \ref{tab:init}. The values presented in this table represent the quantiles 0.5, 0.995 and 0.005 and follow both the geophysical (Sec \ref{sec:results1}) and  thermal (Sec \ref{sec:thermal_constraint}) constraints.}
    \begin{tabular}{c| l c c c c c c }
    Number of profiles & & \multicolumn{1}{c}{Cr} & \multicolumn{1}{c}{Li} & \multicolumn{1}{c}{Ma} & \multicolumn{1}{c}{CMB} & \multicolumn{1}{c}{\af{LC}} & \multicolumn{1}{c}{SIC}\\
    \hline
    & & \multicolumn{6}{c}{\bf{5-layer profiles}} \\
     38,441 &  {\bf{Without BML}}  & & &  & & \\
    &   Radius [km] & 3,389.5 & 3,342$^{3,361}_{3,320}$ & 2,884$^{3,343}_{2,410}$ & 1,785$^{1,899}_{1,702}$ & 1,642$^{1,774}_{1,537}$ & - \\
     &  Density [g.cm$^{-3}$] & 2.90$^{3.10}_{2.70}$ & 3.34$^{3.53}_{3.01}$ & 3.71$^{4.27}_{3.48}$ & 5.14$^{6.78}_{3.64}$ & 6.60$^{7.71}_{5.68}$ & - \\
    &   Viscosity [log$_{10}$(Pa.s)] & - & 21.54$^{29.63}_{17.46}$ & 19.74$^{28.52}_{18.83}$ & 9.82$^{13.84}_{1.72}$ & 10.09$^{29.76}_{1.07}$ & - \\
    & Rigidity [GPa] & 22.45$^{24.99}_{20.04}$ & 45.08$^{69.60}_{20.11}$ & 76.17$^{126.33}_{53.61}$ & - & - & -\\
    \\
     12,050 &  {\bf{With BML}}  & & &  & & \\
    &   Radius [km] & 3,389.5 & 3,345$^{3,360}_{3,320}$ & 3,056$^{3,349}_{2,414}$ & 1,789$^{1,897}_{1,701}$ & 1,653$^{1,776}_{1,540}$ & - \\
     &  Density [g.cm$^{-3}$] & 2.86$^{3.10}_{2.71}$ & 3.27$^{3.52}_{3.05}$ & 3.63$^{4.19}_{3.49}$ & 5.08$^{6.73}_{3.68}$ & 6.56$^{7.86}_{5.75}$ & - \\
    &   Viscosity [log$_{10}$(Pa.s)] & - & 18.70$^{21.65}_{16.69}$ & 20.10$^{22.89}_{19.46}$ & 13.02$^{13.76}_{10.20}$ & 7.64$^{12.38}_{1.22}$ & -\\
    & Rigidity [GPa] & 22.77$^{24.70}_{20.03}$ & 44.31$^{69.01}_{20.62}$ & 73.58$^{108.22}_{52.09}$ & - & - & -\\
    \\
    & & \multicolumn{6}{c}{\bf{6-layer profiles}} \\
    36,893 &  {\bf{Without BML}}  & & &  & & \\
    &   Radius [km] & 3,389.5 & 3,340$^{3,361}_{3,320}$ & 2,882$^{3,335}_{2,410}$ & 1,793$^{1,899}_{1,701}$ & 1,647$^{1,766}_{1,535}$ & 466$^{796}_{1}$\\
     &  Density [g.cm$^{-3}$] & 2.90$^{3.10}_{2.70}$ & 3.34$^{3.52}_{3.02}$ & 3.71$^{4.27}_{3.46}$ & 5.04$^{6.70}_{3.61}$ & 6.54$^{7.67}_{5.42}$ & 8.23$^{9.98}_{6.08}$ \\
    &   Viscosity [log$_{10}$(Pa.s)] & - & 22.82$^{29.82}_{16.63}$ & 20.59$^{29.10}_{18.01}$ & 11.61$^{15.82}_{2.85}$ & 10.78$^{29.71}_{1.15}$ & - \\
    & Rigidity [GPa] & 22.48$^{24.95}_{20.01}$ & 44.25$^{69.86}_{20.35}$ & 74.36$^{125.13}_{21.23}$ & - & - & - \\
    \hline
    \end{tabular}
\label{tab:sol_LS}
\end{table}

\begin{figure}
    \centering
    \includegraphics[scale=0.6]{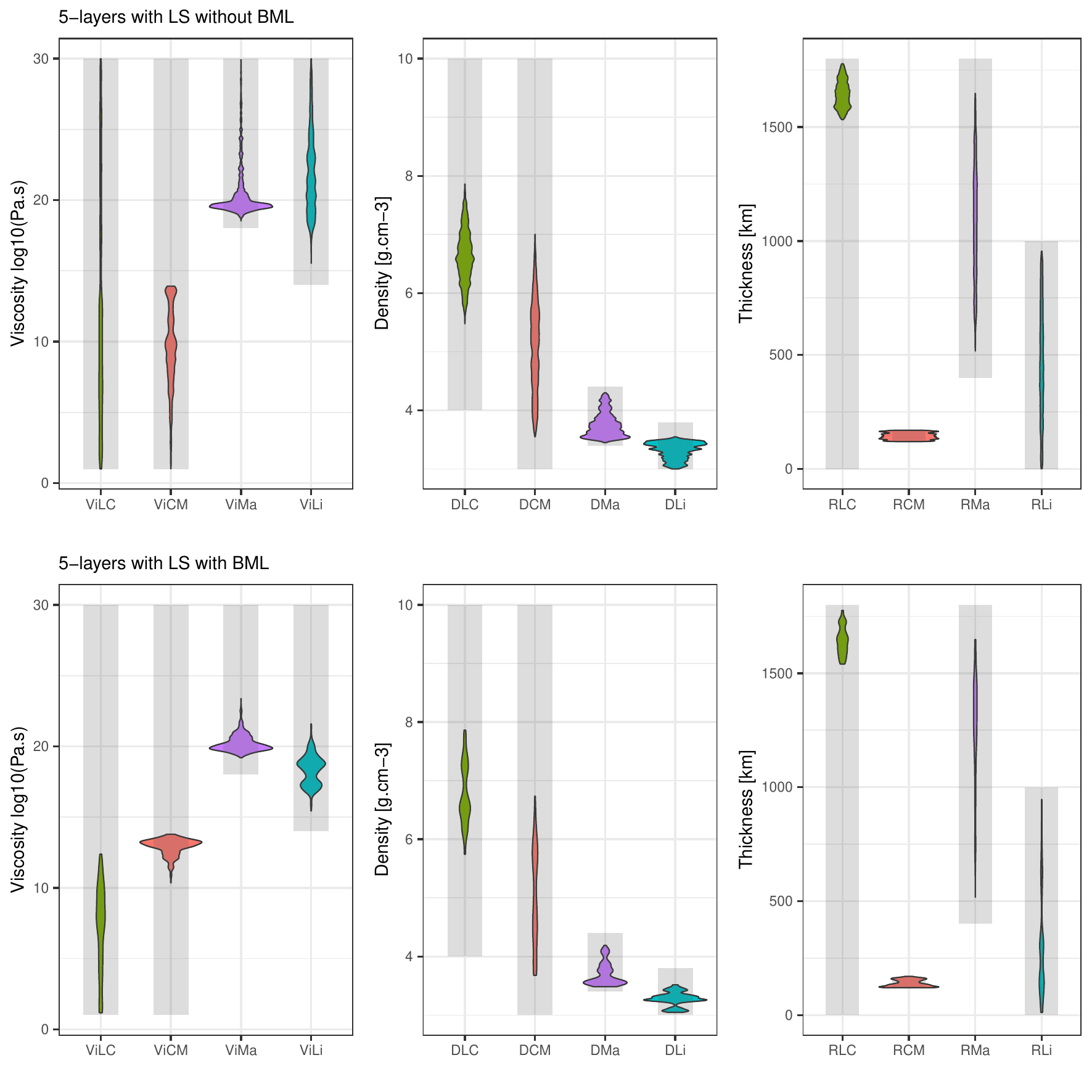}
    \caption{Distributions in viscosity, density and thickness for 5-layer profiles, meeting both geophysical and thermal constraints, without (upper row) and with (lower row) BML. The grey zones picture the exploration priors taken for each of the parameters after filtering with mass and moment of inertia.}
    \label{fig:violon_all}
\end{figure}

\begin{figure}
    \centering
    \includegraphics[width=0.9\textwidth]{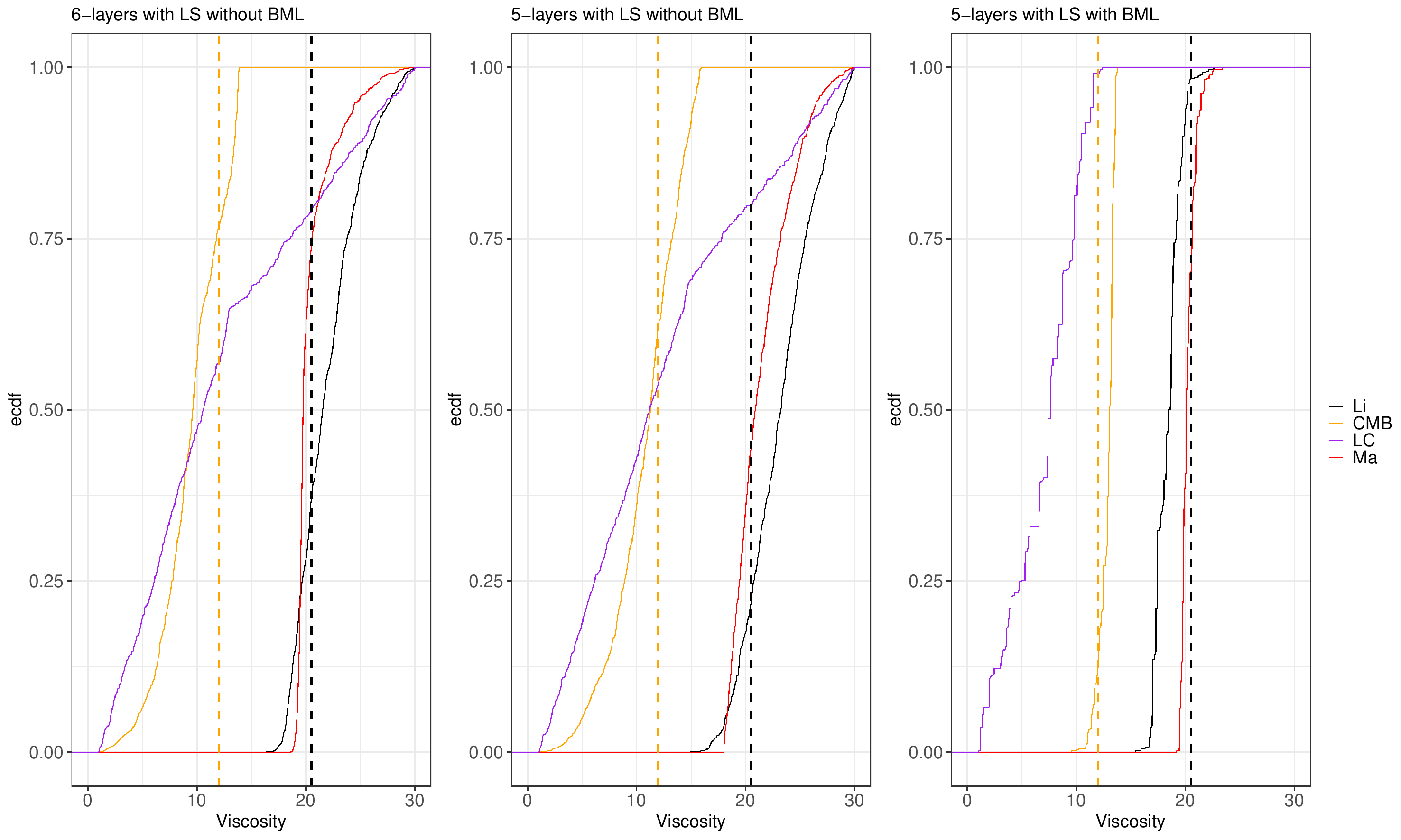}
        \caption{Cumulative histograms of viscosities for 5-layer profiles with and without BML and for 6-layer profiles without BML. Profiles meeting both geophysical and thermal constraints. The lithosphere histogram is given in black, the CMB in orange, the LC in purple and the mantle in red. The dashed line in black is the 0.9 quantile of the lithosphere viscosity where the orange dashed line if the 0.1 quantile of the CMB viscosity.}
    \label{fig:proba}
\end{figure}

\begin{figure}
    \centering
    \includegraphics[width=0.8\textwidth]{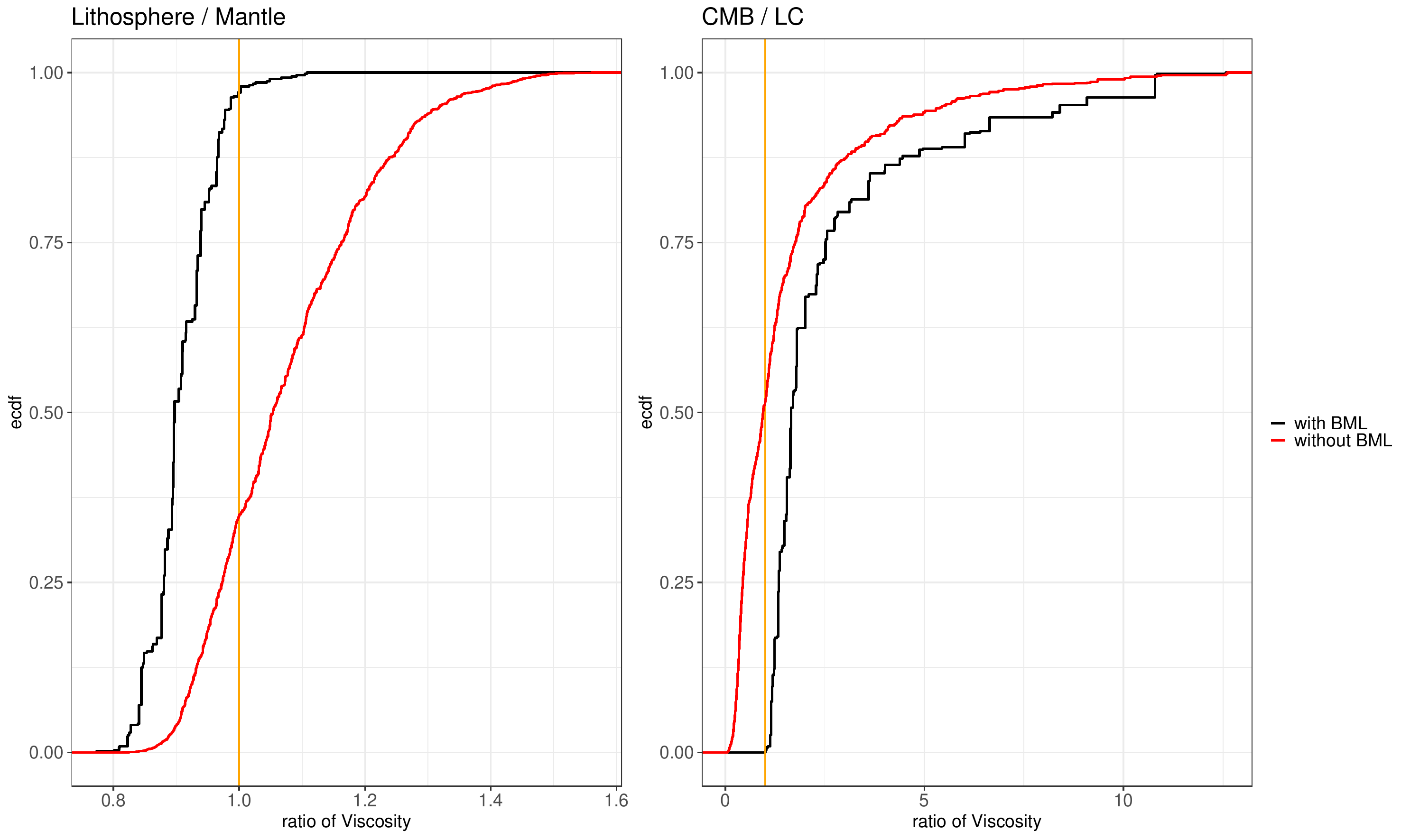}
        \caption{Cumulative histograms of the ratios of viscosity $\eta_{Li}/\eta_{Ma}$ and $\eta_{CMB}/\eta_{LC}$ for 5-layer profiles with (black) and without (red) BML. Profiles meeting both geophysical and thermal constraints. The orange lines indicate $\eta_{Li}/\eta_{Ma} = \eta_{CMB}/\eta_{LC} = 1$ }
    \label{fig:proba_ratio}
\end{figure}

\section{Conclusion}
\label{sec:conclusion}

Despite the very short periods of excitation, tidal deformation is very efficient to constraint the Mars internal structure. New estimations of densities and thicknesses for 5-layer and 6-layer profiles were then obtained and give good agreement with previous studies. Furthermore, combined with thermal constraints, this approach gives also new tools for estimating layer viscosities and thus, monitoring the Mars interior, and more precisely to discuss the existence of a BML.\\
A major result is indeed the impact of BML on the viscosities of the different layers. In fact, the use of different values for the activation enthalpy with and without BML as proposed by \cite{samuel23}, induces very different viscosity distribution for all the layers. 
In our study, the existence of the BML requires that $\eta_{CMB}> 10^{10}$Pa.s, $\eta_{LC}< 10^{13}$Pa.s and $\eta_{Li}< 10^{22}$Pa.s. We also confirm that, with the presence of a BML, the viscosity of Martian lithosphere is, at least, by one order of magnitude lower than the viscosity of mantle as proposed by \cite{khan2023dlayer}. This result has profound implications as it prevents the possibility of a strict Earth-like plate tectonics on Mars and be related to the contrasted degree of hydration between Earth and Mars mantle. 
Without BML, there is no inversion of viscosity between the lithosphere and the mantle and the profiles follow an Earth-like scenario.
We also indicate that the presence of the BML imposes a viscosity of the CMB to be higher that the one of the liquid core. 
Finally, the presence of a BML prevents the existence of a solid core.
{As a putative Martian BML would have multiple consequences both for the lithosphere, the mantle itself and the core structure, geophysical investigations in order to confirm or reject this hypothesis should be a priority. { A best knowledge of viscosity profile would also be of help for this detection.}}

\section{Acknowledgements}
{A.G. thanks EUR Spectrum for excellence grant and A.F and C.G. thank axe transverse "Terrestrial Planets" from OCA for support. }

\appendix

\renewcommand{\thefigure}{A\arabic{figure}}
\setcounter{figure}{0}
\renewcommand{\thetable}{A\arabic{table}}
\setcounter{table}{0}
\section{Results for 6-layer profiles}
\label{sec:appendix}
Are given in this appendix the distribution in viscosity, density and thickness of the 6-layer profiles considering only the geophysical constraints on Fig. \ref{fig:violin_visco_6L} and on Fig. \ref{fig:violon_6LwF}, the same distributions but for profiles selected in also accounting for thermal constraints. As explained in Sec. \ref{sec:nocorewBML}, only the case of 6-layer profiles without BML is presented. By comparisons with Figs. \ref{fig:violin_visco_5L_basic} and \ref{fig:violon_all}, it is interesting to note that the introduction of the solid inner core induces a degradation of the thickness constraints on the liquid core, both with and without the thermal constraints (Figs. \ref{fig:violon_6LwF} and \ref{fig:violin_visco_6L} respectively). This is consistent with a weaker sensitivity of the geophysical constraints to the deeper layers. The fact that the interval of possible densities for the liquid core is similar for the 5-layer and 6-layer profiles is driven by the non-inversion of density between layers imposed to all the profiles and the small volume of the solid inner core (see Sec. \ref{sec:results} for discussion).

\begin{figure}
    \centering
\includegraphics[width=0.7\textwidth]{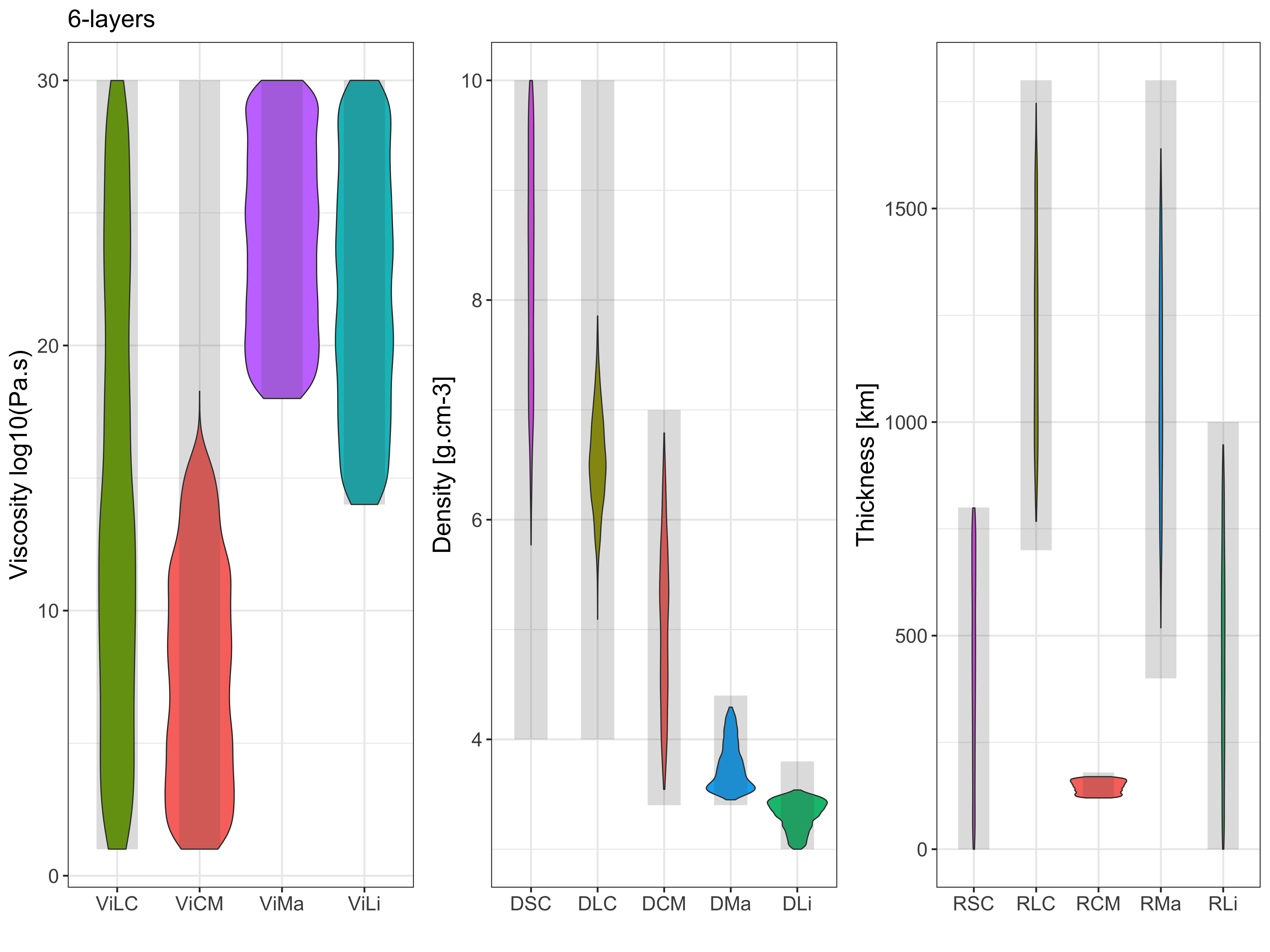}
    \caption{Distribution in viscosity, density and thickness for 6-layer profiles presented in Table \ref{tab:sol_andrade_5&6l}. The gray bars in the background correspond to the exploration intervals of each parameter given in Table \ref{tab:exploration_range} before the $\chi^2$ filtering detailed in Sec. \ref{sec:method}.}
    \label{fig:violin_visco_6L}
\end{figure}

\begin{figure}
    \centering
    \includegraphics[width=0.7\textwidth]{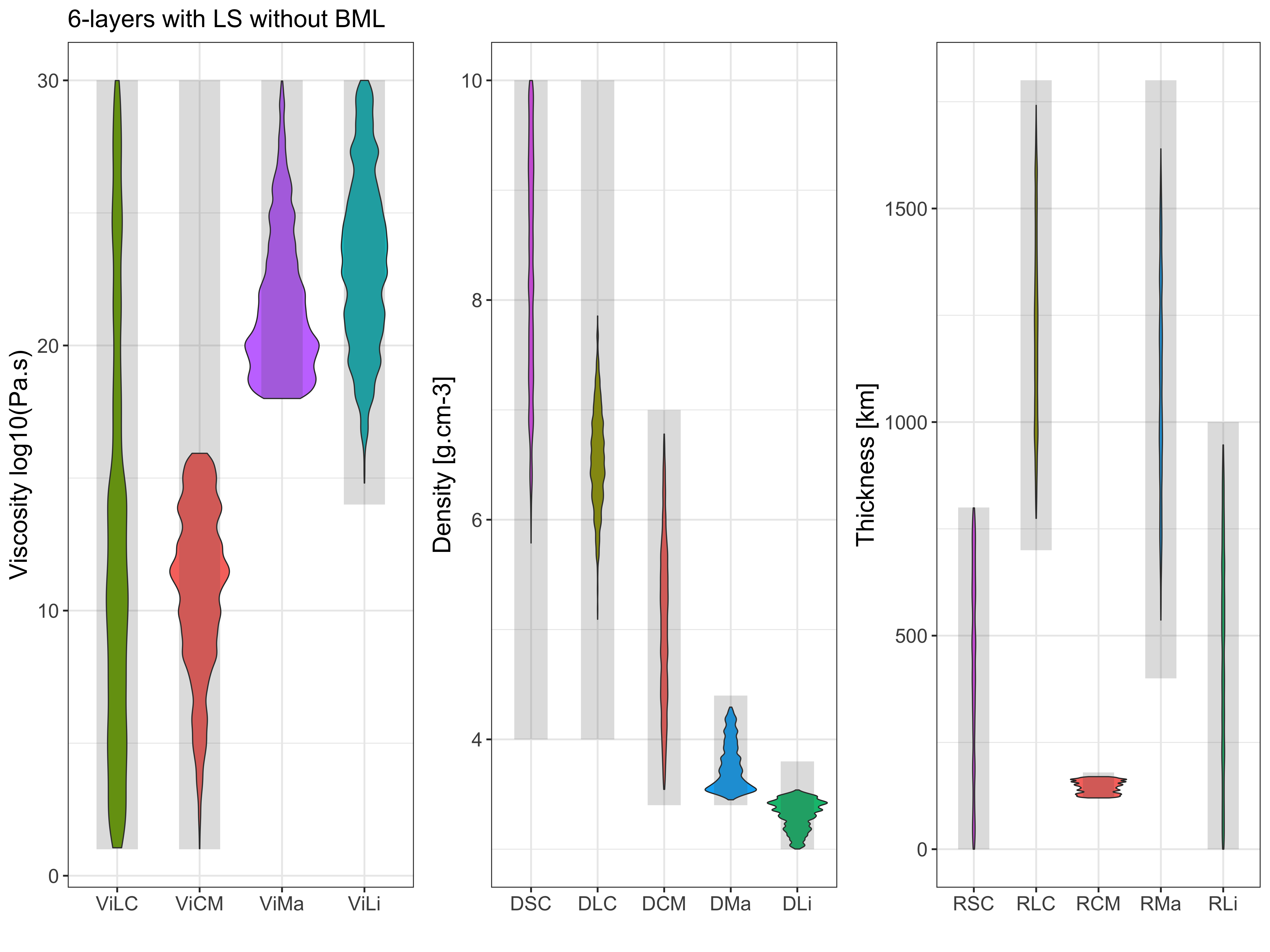}
    \caption{Distributions in viscosity, density and thickness for  the 6-layer profiles without BML, meeting both geophysical and thermal constraints. The intervals are given at 90$\%$ C.L.. The grey zones picture the exploration priors taken for each of the parameters after filtering with mass and moment of inertia.}
    \label{fig:violon_6LwF}
\end{figure}


\bibliographystyle{aasjournal}
\bibliography{Bibliographie.bib}

\begin{thebibliography}{}
\expandafter\ifx\csname natexlab\endcsname\relax\def\natexlab#1{#1}\fi
\providecommand{\url}[1]{\href{#1}{#1}}
\providecommand{\dodoi}[1]{doi:~\href{http://doi.org/#1}{\nolinkurl{#1}}}
\providecommand{\doeprint}[1]{\href{http://ascl.net/#1}{\nolinkurl{http://ascl.net/#1}}}
\providecommand{\doarXiv}[1]{\href{https://arxiv.org/abs/#1}{\nolinkurl{https://arxiv.org/abs/#1}}}

\bibitem[{{Andrault} {et~al.}(2018){Andrault}, {Pesce}, {Manthilake},
  {Monteux}, {Bolfan-Casanova}, {Chantel}, {Novella}, {Guignot}, {King},
  {Iti{\'e}}, \& {Hennet}}]{andrault2018deep}
{Andrault}, D., {Pesce}, G., {Manthilake}, G., {et~al.} 2018, Nature
  Geoscience, 11, 139, \dodoi{10.1038/s41561-017-0053-9}

\bibitem[{Bagheri {et~al.}(2019)Bagheri, Khan, Al-Attar, Crawford, \&
  Giardini}]{bagheri2019tidal}
Bagheri, A., Khan, A., Al-Attar, D., Crawford, O., \& Giardini, D. 2019,
  Journal of Geophysical Research: Planets, 124, 2703

\bibitem[{Bagheri {et~al.}(2022)Bagheri, Efroimsky, Castillo-Rogez, Goossens,
  Plesa, Rambaux, Rhoden, Walterov\`a, Khan, \& Giardini}]{Bagheri22}
Bagheri, A., Efroimsky, M., Castillo-Rogez, J., {et~al.} 2022, Advances in
  Geophysics, 63, 231

\bibitem[{Banerdt {et~al.}(2020)Banerdt, Smrekar, Banfield, Giardini, Golombek,
  Johnson, Lognonn{\'e}, Spiga, Spohn, Perrin, {et~al.}}]{banerdt2020initial}
Banerdt, W.~B., Smrekar, S.~E., Banfield, D., {et~al.} 2020, Nature Geoscience,
  13, 183

\bibitem[{Briaud {et~al.}(2023{\natexlab{a}})Briaud, Ganino, Fienga, M{\'e}min,
  \& Rambaux}]{briaud2023lunar}
Briaud, A., Ganino, C., Fienga, A., M{\'e}min, A., \& Rambaux, N.
  2023{\natexlab{a}}, Nature, 1

\bibitem[{Briaud {et~al.}(2023{\natexlab{b}})Briaud, Fienga, Melini, Rambaux,
  M{\'e}min, Spada, Saliby, Hussmann, Stark, Viswanathan,
  {et~al.}}]{briaud2023constraints}
Briaud, A., Fienga, A., Melini, D., {et~al.} 2023{\natexlab{b}}, Icarus, 115426

\bibitem[{{Doglioni} {et~al.}(2011){Doglioni}, {Ismail-Zadeh}, {Panza}, \&
  {Riguzzi}}]{doglioni2011lithosphere}
{Doglioni}, C., {Ismail-Zadeh}, A., {Panza}, G., \& {Riguzzi}, F. 2011, Physics
  of the Earth and Planetary Interiors, 189, 1,
  \dodoi{10.1016/j.pepi.2011.09.006}

\bibitem[{Drilleau {et~al.}(2022)Drilleau, Samuel, Garcia, Rivoldini, Perrin,
  Michaut, Wieczorek, Tauzin, Connolly, Meyer,
  {et~al.}}]{drilleau2022marsquake}
Drilleau, M., Samuel, H., Garcia, R.~F., {et~al.} 2022, Journal of Geophysical
  Research: Planets, 127, e2021JE007067

\bibitem[{Dur{\'a}n {et~al.}(2022)Dur{\'a}n, Khan, Ceylan, Charalambous, Kim,
  Drilleau, Samuel, \& Giardini}]{duran2022observation}
Dur{\'a}n, C., Khan, A., Ceylan, S., {et~al.} 2022, Geophysical Research
  Letters, 49, e2022GL100887

\bibitem[{Duran {et~al.}(2022)Duran, Khan, Ceylan, Zenh{\"a}usern, Staehler,
  Clinton, \& Giardini}]{duran2022seismology}
Duran, C., Khan, A., Ceylan, S., {et~al.} 2022, Physics of the Earth and
  Planetary Interiors, 325, 106851

\bibitem[{Elkins-Tanton {et~al.}(2003)Elkins-Tanton, Parmentier, \&
  Hess}]{elkins2003magma}
Elkins-Tanton, L.~T., Parmentier, E., \& Hess, P. 2003, Meteoritics \&
  Planetary Science, 38, 1753

\bibitem[{{Helffrich}(2017)}]{helffrich17}
{Helffrich}, G. 2017, Progress in Earth and Planetary Science, 4, 24,
  \dodoi{10.1186/s40645-017-0139-4}

\bibitem[{Hemingway \& Driscoll(2021)}]{hemingway2021history}
Hemingway, D.~J., \& Driscoll, P.~E. 2021, Journal of Geophysical Research:
  Planets, 126, e2020JE006663

\bibitem[{Horleston {et~al.}(2022)Horleston, Clinton, Ceylan, Giardini,
  Charalambous, Irving, Lognonn{\'e}, St{\"a}hler, Zenh{\"a}usern, Dahmen,
  {et~al.}}]{horleston2022far}
Horleston, A.~C., Clinton, J.~F., Ceylan, S., {et~al.} 2022, The Seismic
  Record, 2, 88

\bibitem[{Huang {et~al.}(2022)Huang, Schmerr, King, Kim, Rivoldini, Plesa,
  Samuel, Maguire, Karakostas, Leki{\'c}, {et~al.}}]{huang2022seismic}
Huang, Q., Schmerr, N.~C., King, S.~D., {et~al.} 2022, Proceedings of the
  National Academy of Sciences, 119, e2204474119

\bibitem[{Irving {et~al.}(2023)Irving, Leki{\'c}, Dur{\'a}n, Drilleau, Kim,
  Rivoldini, Khan, Samuel, Antonangeli, Banerdt, {et~al.}}]{irving2023first}
Irving, J.~C., Leki{\'c}, V., Dur{\'a}n, C., {et~al.} 2023, Proceedings of the
  National Academy of Sciences, 120, e2217090120

\bibitem[{{Karato} {et~al.}(2008){Karato}, {Jung}, {Katayama}, \&
  {Skemer}}]{karato2088geodynamic}
{Karato}, S.-I., {Jung}, H., {Katayama}, I., \& {Skemer}, P. 2008, Annual
  Review of Earth and Planetary Sciences, 36, 59,
  \dodoi{10.1146/annurev.earth.36.031207.124120}

\bibitem[{{Khan} {et~al.}(2023){Khan}, {Huang}, {Duran}, {Sossi}, {Murakami},
  \& {Giardini}}]{khan2023dlayer}
{Khan}, A., {Huang}, D., {Duran}, C., {et~al.} 2023, in LPI Contributions, Vol.
  2806, LPI Contributions, 1448

\bibitem[{Khan {et~al.}(2023)Khan, Huang, Dur{\'a}n, Sossi, Giardini, \&
  Murakami}]{khan2023evidence}
Khan, A., Huang, D., Dur{\'a}n, C., {et~al.} 2023, Nature, 622, 718

\bibitem[{{Khan} {et~al.}(2018){Khan}, {Liebske}, {Rozel}, {Rivoldini},
  {Nimmo}, {Connolly}, {Plesa}, \& {Giardini}}]{khan2018geophysical}
{Khan}, A., {Liebske}, C., {Rozel}, A., {et~al.} 2018, Journal of Geophysical
  Research (Planets), 123, 575, \dodoi{10.1002/2017JE005371}

\bibitem[{Khan {et~al.}(2021)Khan, Ceylan, van Driel, Giardini, Lognonn{\'e},
  Samuel, Schmerr, St{\"a}hler, Duran, Huang, {et~al.}}]{khan2021upper}
Khan, A., Ceylan, S., van Driel, M., {et~al.} 2021, Science, 373, 434

\bibitem[{Kim {et~al.}(2023)Kim, Duran, Giardini, Plesa, St{\"a}hler, Boehm,
  Lekic, McLennan, Ceylan, Clinton, {et~al.}}]{kim2023global}
Kim, D., Duran, C., Giardini, D., {et~al.} 2023

\bibitem[{Konopliv {et~al.}(2016)Konopliv, Park, \&
  Folkner}]{konopliv2016improved}
Konopliv, A.~S., Park, R.~S., \& Folkner, W.~M. 2016, Icarus, 274, 253

\bibitem[{Konopliv {et~al.}(2020)Konopliv, Park, Rivoldini, Baland, Le~Maistre,
  Van~Hoolst, Yseboodt, \& Dehant}]{konopliv2020detection}
Konopliv, A.~S., Park, R.~S., Rivoldini, A., {et~al.} 2020, Geophysical
  Research Letters, 47, e2020GL090568

\bibitem[{Li {et~al.}(2022)Li, Beghein, McLennan, Horleston, Charalambous,
  Huang, Zenh{\"a}usern, Bozda{\u{g}}, Pike, Golombek,
  {et~al.}}]{li2022constraints}
Li, J., Beghein, C., McLennan, S.~M., {et~al.} 2022, Nature Communications, 13,
  7950

\bibitem[{Lillis {et~al.}(2005)Lillis, Manga, Mitchell, Lin, \&
  Acu{\~n}a}]{lillis2005evidence}
Lillis, R.~J., Manga, M., Mitchell, D.~L., Lin, R.~P., \& Acu{\~n}a, M.~H.
  2005, Lunar and Planetary Science XXXVI, Part 12

\bibitem[{{Melini} {et~al.}(2022){Melini}, {Saliby}, \&
  {Spada}}]{2022GeoJI.231.1502M}
{Melini}, D., {Saliby}, C., \& {Spada}, G. 2022, Geophysical Journal
  International, 231, 1502, \dodoi{10.1093/gji/ggac263}

\bibitem[{Plesa {et~al.}(2021)Plesa, Bozda{\u{g}}, Rivoldini, Knapmeyer,
  McLennan, Padovan, Tosi, Breuer, Peter, Staehler,
  {et~al.}}]{plesa2021seismic}
Plesa, A.-C., Bozda{\u{g}}, E., Rivoldini, A., {et~al.} 2021, Journal of
  Geophysical Research: Planets, 126, e2020JE006755

\bibitem[{Post(1930)}]{post1930generalized}
Post, E.~L. 1930, Transactions of the American Mathematical Society, 32, 723

\bibitem[{Pou {et~al.}(2022)Pou, Nimmo, Rivoldini, Khan, Bagheri, Gray, Samuel,
  Lognonn{\'e}, Plesa, Gudkova, {et~al.}}]{pou2022tidal}
Pou, L., Nimmo, F., Rivoldini, A., {et~al.} 2022, Journal of Geophysical
  Research: Planets, 127, e2022JE007291

\bibitem[{Rivoldini {et~al.}(2011)Rivoldini, Van~Hoolst, Verhoeven, Mocquet, \&
  Dehant}]{rivoldini2011geodesy}
Rivoldini, A., Van~Hoolst, T., Verhoeven, O., Mocquet, A., \& Dehant, V. 2011,
  Icarus, 213, 451

\bibitem[{{Saliby} {et~al.}(2023){Saliby}, {Fienga}, {Briaud}, {M{\'e}min}, \&
  {Herrera}}]{Saliby2023}
{Saliby}, C., {Fienga}, A., {Briaud}, A., {M{\'e}min}, A., \& {Herrera}, C.
  2023, Planetary and Space Sciences, 231, 105677,
  \dodoi{10.1016/j.pss.2023.105677}

\bibitem[{Samuel {et~al.}(2022)Samuel, Drilleau, Garcia, Huang, Rivoldini,
  Lognonn{\'e}, \& Banerdt}]{samuel2022testing}
Samuel, H., Drilleau, M., Garcia, R., {et~al.} 2022, Testing the Presence Deep
  Martian Mantle Layering in the light of InSight Seismic Data, Tech. rep.,
  Copernicus Meetings

\bibitem[{{Samuel} {et~al.}(2023){Samuel}, {Drilleau}, {Rivoldini}, {Xu},
  {Huang}, {Garcia}, {Leki{\'c}}, {Irving}, {Badro}, {Lognonn{\'e}},
  {Connolly}, {Kawamura}, {Gudkova}, \& {Banerdt}}]{samuel23}
{Samuel}, H., {Drilleau}, M., {Rivoldini}, A., {et~al.} 2023, \nat, 622, 712,
  \dodoi{10.1038/s41586-023-06601-8}

\bibitem[{Seidelmann {et~al.}(2002)Seidelmann, Abalakin, Bursa, Davies,
  de~Bergh, Lieske, Oberst, Simon, Standish, Stooke, \&
  Thomas}]{seidelmann2002}
Seidelmann, P.~K., Abalakin, V.~K., Bursa, M., {et~al.} 2002, Celestial
  Mechanics and Dynamical Astronomy, 82, 83, \dodoi{10.1023/A:1013939327465}

\bibitem[{Spada(2008)}]{spada2008alma}
Spada, G. 2008, Computers \& Geosciences, 34, 667

\bibitem[{Spada \& Boschi(2006)}]{spada2006using}
Spada, G., \& Boschi, L. 2006, Geophysical Journal International, 166, 309

\bibitem[{St{\"a}hler {et~al.}(2021)St{\"a}hler, Khan, Ceylan, Duran, Garcia,
  Giardini, Huang, Kim, Lognonn{\'e}, Maguire, {et~al.}}]{stahler2021seismic}
St{\"a}hler, S.~C., Khan, A., Ceylan, S., {et~al.} 2021, in EGU General
  Assembly Conference Abstracts, EGU21--13310

\bibitem[{{Wanke} \& {Dreibus}(1994)}]{wanke1994chemistry}
{Wanke}, H., \& {Dreibus}, G. 1994, Philosophical Transactions of the Royal
  Society of London Series A, 349, 285, \dodoi{10.1098/rsta.1994.0132}

\bibitem[{Widder(1941)}]{widder1941laplace}
Widder, D. 1941, Princeton Mathematical Series

\bibitem[{Widder(1934)}]{widder1934inversion}
Widder, D.~V. 1934, Transactions of the American Mathematical Society, 36, 107

\bibitem[{Wieczorek {et~al.}(2022)Wieczorek, Broquet, Mclennan, Rivoldini,
  Golombek, Antonangeli, Beghein, Giardini, Gudkova, Gyalay,
  {et~al.}}]{wieczorek2022insight}
Wieczorek, M.~A., Broquet, A., Mclennan, S.~M., {et~al.} 2022, Journal of
  Geophysical Research: Planets, 127, e2022JE007298

\bibitem[{Zeff \& Williams(2019)}]{zeff2019fractional}
Zeff, G., \& Williams, Q. 2019, Geophysical Research Letters, 46, 10997

\end{thebibliography}

\end{document}